\documentclass[runningheads]{llncs}
\usepackage[T1]{fontenc}
\usepackage{graphicx}
\usepackage[ruled,vlined,linesnumbered,lined,boxed,commentsnumbered]{algorithm2e}
\usepackage{subfig}
\usepackage{multirow}
\usepackage{caption}
\usepackage{subfig}

\usepackage{comment}

\begin{document}

\title{Whispy: Adapting STT Whisper Models to Real-Time Environments}
\titlerunning{Whispy: Real-Time Whisper transcriptions}

\author{Antonio Bevilacqua\inst{1} \and Paolo Saviano\inst{1} \and
Alessandro Amirante\inst{1,2} \and Simon Pietro Romano \inst{1,2}}
\authorrunning{A. Bevilacqua et al.}

\institute{Meetecho LTD, Napoli, Italy\\
\url{https://www.meetecho.com/en} \and
University of Napoli Federico II\\
\url{https://www.unina.it}}

\maketitle

\begin{abstract}
Large general-purpose transformer models have recently become the mainstay in the realm of speech analysis. In particular, Whisper achieves state-of-the-art results in relevant tasks such as speech recognition, translation, language identification, and voice activity detection. However, Whisper models are not designed to be used in real-time conditions, and this limitation makes them unsuitable for a vast plethora of practical applications. In this paper, we introduce Whispy, a system intended to bring live capabilities to the Whisper pretrained models. As a result of a number of architectural optimisations, Whispy is able to consume live audio streams and generate high level, coherent voice transcriptions, while still maintaining a low computational cost. We evaluate the performance of our system on a large repository of publicly available speech datasets, investigating how the transcription mechanism introduced by Whispy impacts on the Whisper output. Experimental results show how Whispy excels in robustness, promptness, and accuracy.

\keywords{speech-to-text \and whisper \and transcription \and real-time}
\end{abstract}

\section{Introduction}

In recent years, Automatic Speech Recognition (ASR) systems have gained momentum across diverse sectors, propelled by advancements in machine learning, deep neural networks, and natural language processing techniques \cite{Prabhavalkar_2024,Zimmermann_2024}. These systems, equipped with sophisticated algorithms, have become indispensable tools for a wide range of applications, spanning from virtual assistants and smart home devices to customer service automation and healthcare diagnostics. The surge in ASR adoption is fueled by the proliferation of commercial solutions offered by leading tech companies and startups alike~\cite{Zimmermann_2024}. These solutions, often accessible as paid services through web-based Application Programming Interfaces (APIs), provide developers with convenient and scalable access to powerful speech recognition capabilities without the need for extensive expertise in machine learning or signal processing. As a result, businesses and developers can seamlessly integrate ASR functionality into their applications, enabling end-to-end speech-to-text transcription, voice commands, sentiment analysis, and more~\cite{Prabhavalkar_2024}.

Lately, the family of large-scale models called Whisper \cite{radford2022robust}, released by the company OpenAI, secured state-of-the-art performance in the realm of ASR. Whisper models are multilingual, multitask systems capable of achieving human accuracy in transcribing and translating recorded audio, identifying spoken languages, and detecting voice activity (Voice Activity Detection, or VAD). One of the key strengths of Whisper models lies in their ability to achieve human-level accuracy in transcribing and translating recorded audio. This means that they can accurately convert spoken words into text and even translate them into different languages with a level of precision that rivals human capabilities. Additionally, Whisper models excel in identifying the languages being spoken and detecting voice activity within audio recordings. These capabilities have significant implications for various applications, including speech-to-text transcription services, multilingual communication platforms, and voice-controlled systems. By leveraging the power of large-scale deep learning models, Whisper has pushed the boundaries of what is possible in ASR, offering unprecedented accuracy and versatility in processing and understanding spoken language. Yet, Whisper models are limited to offline use only. Whisper requires the target speech data to be fully available at inference time, and generates transcriptions or translations at a speed that depends on many factors and is not easy to determine for long audio sources. This highly hinders their potential to be adopted in real-time applications, particularly in contexts where low-latency speech analysis and ASR are essential, such as in web conferencing.

In this paper, we present Whispy, a novel adaptation structure around Whisper models that allows for live transcription of audio streams. Whispy is designed as a self-contained ASR service, capable of consuming real-time speech data and producing precise and reliable transcriptions at a reduced computational cost and with an acceptable temporal delay. This is achieved by processing short audio chunks that accumulate within a shifting buffer. A straightforward and effective agreement algorithm, based on the Levenshtein distance~\cite{editdistance} between strings, extracts the most accurate transcript suggestions when overlapping portions of the buffer are transcribed. Whispy is highly configurable and flexible, and performed consistently well across all of our experimental campaigns.

In real-time communication and collaboration contexts, ASR techniques play a pivotal role by enabling real-time speech transcription, thereby enhancing these environments with functionalities such as speech summarisation and diarisation. These capabilities offer users the ability to work with spoken content in real-time, facilitating tasks like extracting key insights from conversations and identifying speakers. The Whispy system, specifically designed with this purpose in mind, is intended to seamlessly integrate with various real-time communication systems. While Whispy's functionality remains agnostic to the underlying real-time communication platform, the current implementation discussed in this paper assumes integration with a WebRTC-enabled architecture, leveraging the open-source Janus media server developed by Meetecho at its core~\cite{IPTComm2014-Janus}.

The rest of this paper is organised as follows. Section~\ref{section:related} briefly explores the space of currently available solutions for live ASR. In Section~\ref{section:methodology} we present the Whispy architecture and working principles. Experimental setups and results are discussed in Section~\ref{section:results}. Finally, a compendium of our contribution, with some closing remarks and suggestions for future extensions, is provided in Section~\ref{section:conclusions}.

\section{Related works}
\label{section:related}

A substantial body of literature has explored capabilities and applications of ASR systems, mainly in offline setups. Spanning from the medical and healthcare context \cite{Huh2023Improving,vargas2024}, to education \cite{10.1145/3308561.3353772,9411528}, surveillance and security \cite{9519395} and smart homes \cite{s23135784}, speech-to-text models have been adopted for a variety of tasks. However, research in this sense has mainly been focusing on improving transcription performance and context-awareness of ASR systems, rather than extending their capabilities to real-time scenarios. In this regard, to the best of our knowledge, most of the currently available solutions come from private companies and startups that offer such services to a price and therefore do not disclose useful insight on their processes.

In the open source domain, interesting projects exist that investigate potential adaptations of Whisper as a real-time Speech-To-Text (STT) system. Within solutions like \texttt{VoiceStreamAI}~\cite{Saccoia2024alesaccoia} or \texttt{Whisper-live}~\cite{Sheikholmolouki2023Alireza29675}, the streaming audio data is split into smaller chunks, and then each chunk is processed using a Whisper model instance. This method has the drawback of not providing Whisper with enough context information, so the transcriptions for short chunks may be of extremely poor quality, in particular when sentences or words are cut at the beginning or the end of a chunk. Similarly, \texttt{Whisper-streaming}~\cite{machacek-dabre-bojar:2023:ijcnlp} offers a client-server solution based on chunking input audio streams and then running a local agreement algorithm to align current transcriptions with the previously transcribed text. Whilst quite similar to our approach in terms of audio management, \texttt{Whisper-streaming} relies on word-level timestamps to extract text updates from transcription sequences of already processed text.

In this paper, we address the main issues that arise from the discussed solutions, and we test our system against commonly used ASR benchmark datasets in order to evaluate both the quality of the produced transcriptions, and the introduced temporal delay.

\section{Data and methodology}
\label{section:methodology}
The purpose of this paper is to evaluate how Whispy, our wrapper designed around the Whisper pretrained models, can be successfully used to transcribe audio data fetched from live sources.

We designed Whispy to be a production-ready system that can be used as a self-contained transcription service. Our objective was twofold: to provide a robust real-time speech-to-text container, and to engineer a flexible multimedia-oriented blueprint for a software structure capable of processing live audio and video streams and feeding them into extensible AI components.

\begin{figure}[h!]
    \centering
    \includegraphics[width=\textwidth]{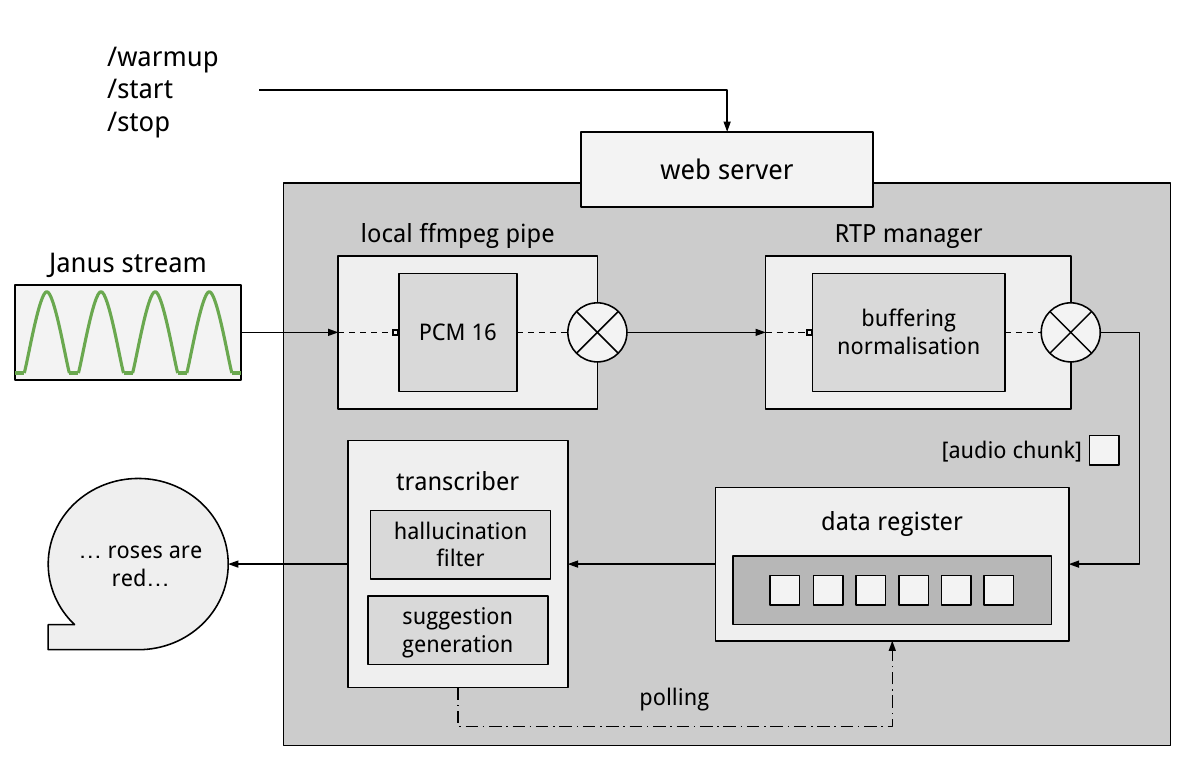}
    \caption{General Whispy service architecture. The overall system lives within an HTTP server, that we use as interface to set up the incoming stream, define the transcription destination, and update options such as model size or Voice Activity Detection (VAD) parameters.}
    \label{fig:architecture}
\end{figure}

As shown in Figure~\ref{fig:architecture}, the main actors in the Whispy transcriber are:

\begin{itemize}
    \item an input pipeline, responsible for managing inbound real-time audio streams and adapting them to formats suitable for any following components;
    \item a shifting data register, in charge of storing the received audio data and making them available to the transcriber or any other entity that may need them;
    \item a transcriber, holding and invoking the actual Whisper model.
\end{itemize}

Furthermore, Whispy exposes a number of web APIs that can be used as signalling entry points to negotiate the connection parameters of input streams and output destinations, as well as start and stop the transcription process.

\subsection{Input pipeline}
Whispy receives and preprocesses incoming data streams through a two-layer pipeline. First, an \texttt{ffmpeg} asynchronous process (created with the use of the \texttt{ffmpeg-python} library~\cite{ffmpegpython}) is configured with all the relevant multimedia information received upon a remote \texttt{/warmup} POST call, such as input codec, number of channels, and audio sampling rate. The \texttt{ffmpeg} process then produces a local Real-time Transfer Protocol (RTP) stream, that is consumed by an RTP client where buffering and further data manipulations may take place.

This input structure presents many advantages. It allows us to fully leverage \texttt{ffmpeg} capabilities for the heavy-lifting of data manipulation, such as encoding and decoding, and it also makes the whole application easily extensible to multiple input/output scenarios. Moreover, it completely decouples the data source from the Whispy service, making it agnostic to the application context.

\subsection{Data register}
\label{methodology:dataregister}
Once the RTP client buffers a set amount of data, it sends it to a data structure that acts as a First In, First Out (FIFO) shifting register. This register can hold an audio waveform up to a fixed length, determined by the formula $C_n * C_d * f_s$, where:

\begin{itemize}
    \item $C_n$ is the number of audio chunks that can be stored in the register;
    \item $C_d$ is the temporal duration in seconds of each audio chunk;
    \item $f_s$ is the incoming audio sampling rate (as produced by the RTP client).
\end{itemize}

As an example, a register that contains $20$ total seconds of audio sampled at $16$ kHz, split over $5$ chunks of $4$ seconds each, will have a length of $5 * 4 * 16000 = 320k$ data points. In addition, the data register keeps track of the absolute number of data blocks that were appended to it, as well as the temporal starting and ending coordinates of its content at any given time.

\subsection{Transcriber}
The transcriber is the component that holds and invokes the actual Whisper model. In our implementation, we use \texttt{faster-whisper}~\cite{githubGitHubSYSTRANfasterwhisper}, which provides inference models quantized and optimized with \texttt{CTranslate2}~\cite{ctranslate2} in order to improve the models' inference performance and considerably reduce their memory requirements.

When a transcription is in progress, the transcriber periodically polls the data register for new audio chunks, using the absolute sequence number as reference. Once a new audio chunk is available, the transcriber fetches a copy of the entire register content and:

\begin{enumerate}
    \item checks the waveform corresponding to the latest audio chunk only for voice activity (pre-VAD). If no voice is detected, the transcriber does not perform any transcription, marks the latest chunk as silent, and flushes the content of the data register;
    \item if the waveform corresponding to the latest audio chunk contains voice activity, the entire waveform is transcribed;
    \item the newly produced transcription is first filtered to detect potential \emph{hallucinations}~\cite{koenecke2024careless}, then processed against the previous transcription to generate a suggestion that fixes the overlap between subsequent transcription strings.
\end{enumerate}

A more detailed explanation of the transcription process is provided in Algorithm~\ref{alg:transcription}. It is important to highlight that VAD operations heavily rely on the pretrained Silero VAD models~\cite{SileroVAD}, which are integrated into the \texttt{faster-whisper} library. The inclusion of a pre-VAD step serves as a strategic optimization, aimed at preventing unnecessary re-transcriptions triggered by periods of silence or non-voice audio segments added to the buffer. Despite the introduction of this pre-VAD step, VAD remains a crucial component of the \texttt{faster-whisper} transcription pipeline. This is because conducting VAD during transcription significantly enhances the quality of the text output and effectively reduces the likelihood of encountering hallucinations or other undesirable artifacts in the transcribed content. By incorporating the pretrained Silero VAD models within the \texttt{faster-whisper} framework, the transcription process benefits from enhanced accuracy and efficiency. This integrated approach ensures a smoother and more reliable transcription experience for users, ultimately improving the overall performance and usability of the system.

Another crucial aspect of our transcription policy is the re-transcription of portions of audio that were already previously processed, as shown on line \ref{alg:trx} in Algorithm \ref{alg:transcription}. This behaviour results in transcript outputs that are context aware, thus eliminating any issues caused by chunks cutting in mid-sentence or mid-word. Moreover, unexpected delays in the pipeline that may result in more than one chunk to be appended to data register without the transcriber fetching them via polling are ultimately resolved without data loss.

\begin{algorithm}
    \SetKwInOut{Input}{Input}
    \SetKwInOut{Output}{Output}
    
    \SetKwData{data}{data}
    \SetKwData{register}{register}
    \SetKwData{model}{model}
    \SetKwData{trx}{trx}
    \SetKwData{lasttrx}{last\_trx}
    \SetKwData{lastchunk}{last\_chunk}
    \SetKwData{hasvoice}{has\_voice\_activity}
    \SetKwData{hallucination}{hallucination}

    \SetKwFunction{vad}{run\_vad}
    \SetKwFunction{flush}{flush}
    \SetKwFunction{getlastchunk}{get\_last\_chunk}
    \SetKwFunction{transcribe}{transcribe}
    \SetKwFunction{ishall}{is\_hallucination}
    \SetKwFunction{fix}{filter\_hallucination}
    \SetKwFunction{suggest}{generate\_suggestion}

    \Input{Shifting register \register with \data content}
    \Input{Transcription model \model}
    \Input{Previous transcription result \lasttrx}
    \Output{Transcription object \trx}
    \BlankLine

    \lastchunk $\leftarrow$ \register.\data.\getlastchunk{}\\
    \hasvoice $\leftarrow$ \vad{\lastchunk}\\
    \BlankLine
    \If{\hasvoice is False}
    {
        \register.\flush{}\\
        \Return{'silence'}
    }

    \trx $\leftarrow$ \model.\transcribe{\data}\label{alg:trx}\\
    \hallucination $\leftarrow$ \ishall{\trx}\\
    \BlankLine

    \If{\hallucination is True}
    {
        \trx $\leftarrow$ \fix{\trx}
    }

    \trx $\leftarrow$ \suggest{\trx, \lasttrx}\\
    \Return{\trx}

    \caption{Transcription algorithm}
    \label{alg:transcription}
\end{algorithm}

Two other important components of the transcription process are the hallucination filter and the suggestion generator. Whilst the first is required because of the intrinsic tendency of large text-generation models to produce, from time to time, unreliable or unpredictable outputs \cite{Ji2023}, we need the latter as Whispy transcribes portions of overlapping audio between pairs of consecutive audio chunks.

\subsubsection{Generating suggestions}

\begin{figure}
    \centering
    \includegraphics[width=\textwidth]{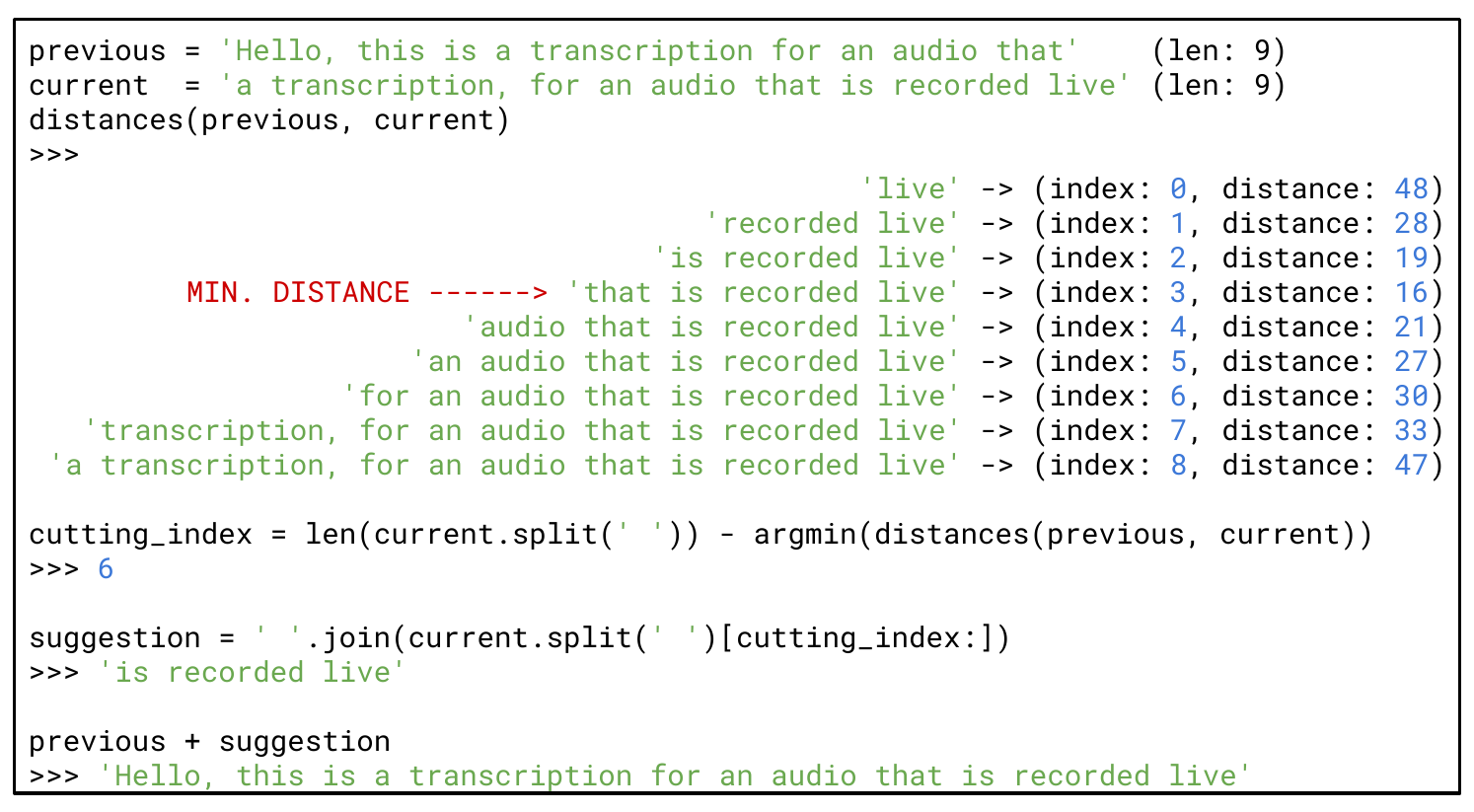}
    \caption{Practical example of the Whispy suggestion mechanism.}
    \label{fig:suggestion}
\end{figure}

When any given audio chunk $n$ is added to the data register and processed for transcription, the produced text will include a transcription of a substantial portion of audio that was already transcribed up to chunk $n - 1$. In order to determine which part of the newly generated text should be used as transcription for the current chunk, we designed a search technique based on the Levenshtein distance~\cite{editdistance}. The Levenshtein distance, often referred to as edit distance, is a simple metric that quantifies the distance between a pair of sequences in terms of how many insertions, deletions and substitutions are required to change one sequence into the other. As shown in Figure~\ref{fig:suggestion}, Whispy generates a sequence of tokens from the transcription of an audio chunk, and then, going backwards from the last token, computes the edit distance between all the subsequences and the transcription text of the previous chunk. The subsequence with the minimum edit distance to the previous transcription text is used to extract the suggested transcription for the current chunk.

\subsubsection{Filtering hallucinations}

Hallucinations occasionally produced by the Whisper models have the potential to extensively disrupt the Whispy transcription outcome. Other than content-oriented hallucinations~\cite{koenecke2024careless}, Whisper can produce hallucinatory artifacts resulting in the repetition of a single token or sequence of tokens~\cite{frieske2024hallucinations}. In our implementation, we designed a naive hallucination filter capable of reliably detecting repeating tokens and skim them before suggestions are produced. As we will discuss in Chapter \ref{section:results}, however, during our test campaigns we did not experience a sufficiently large number of hallucinatory events to determine the goodness of the hallucination filter.

\section{Experimental results}
\label{section:results}
In our experimental campaigns, we tested Whispy with emphasis on transcription quality and live transcription delay. We also compared the results obtained with Whispy against the outcomes of the Whisper offline transcriptions, in order to determine whether the buffering mechanism we designed results in less reliable transcripts, and whether any differences are statistically significant or hold any practical relevance. In doing so, we selected a number of heterogeneous datasets commonly used to test ASR systems, namely:

\begin{itemize}
    \item \textbf{ESIC}~\cite{machacek21_interspeech}, a corpus of recorded speeches from European Parliament sessions, for which human transcriptions are available. ESIC includes \texttt{dev}, \texttt{dev2} and \texttt{test} partitions. We incporporated the latter in our experiments, but we had to exclude a number of items from the partition as they either were incorrectly labeled as containing English speech or carried non-English transcripts.
    \item \textbf{librispeech}~\cite{meister2023librispeechpc}, a dataset of readings from audiobooks for which the text is available. Librispeech is split into a \texttt{clean} partition and a \texttt{other} partition, depending on the audio quality. Out of the box, these datasets come in the form of short audio clips where each clip contains a sentence extracted from a particular book. We merged together clips from the same readings in order to obtain longer audio data.
    \item \textbf{TEDlium}~\cite{Hernandez_2018}, a corpus of recorded TED talks with their transcripts. We tested the legacy \texttt{test} partition, commonly used as benchmark in literature, composed of $11$ talks.
    \item \textbf{Rev16}~\cite{revSpeechText}, a set of 16 recordings from the Rev.ai podcast.
\end{itemize}

We preprocessed all the audio data at our disposal by appending a short period of silence at the end of every clip. This silence ensures that Whispy correctly processes the content of the buffer at the end of the stream whenever the last chunk in it is shorter than the set chunk length. According to the definitions outlined in Section~\ref{methodology:dataregister}, we set a chunk length of $4$ seconds and a buffer length of $5$ chunks, for a total of $20$ seconds of audio. Hence we added $4$ seconds of silence at the end of every tested audio clip. When testing different hyperparameter configurations, and specifically different chunk lengths, we modified the trailing silence accordingly. We designed a flexible testing pipeline that dynamically instantiates all the required services for both Whispy and Whisper systems, then streams all the audio clips within a dataset in order to obtain the Whispy transcripts, invokes the Whisper offline model, and eventually computes all the required metrics based on the ground truth transcription. Both Whispy and Whisper were tested with the model instances \texttt{base}, \texttt{small}, \texttt{medium}, and \texttt{large-v3}, and always configured with the same set of parameters. Our testing pipeline runs on a cloud infrastructure powered by four Tesla T4 cards that are used to run multiple tests in parallel, so each Whispy instance is allocated a single T4 card at any given time.

\subsection{Transcription quality}
We evaluate the transcription quality by computing the Word Error Rate (WER) between any transcription generated by either Whisper or Whispy, and the corresponding ground truth text. WER is a metric based on the edit distance between sequences, that is commonly regarded as a standard when assessing the performance of ASR systems \cite{von_Neumann_2023}. In order to avoid unfair penalisation of hypothesis texts, we first normalise all generated text data by removing punctuation and multiple whitespaces, converting all tokens to lower case, and expanding English contractions.

\begin{figure}[ht!]
    \centering
    \includegraphics[width=\textwidth]{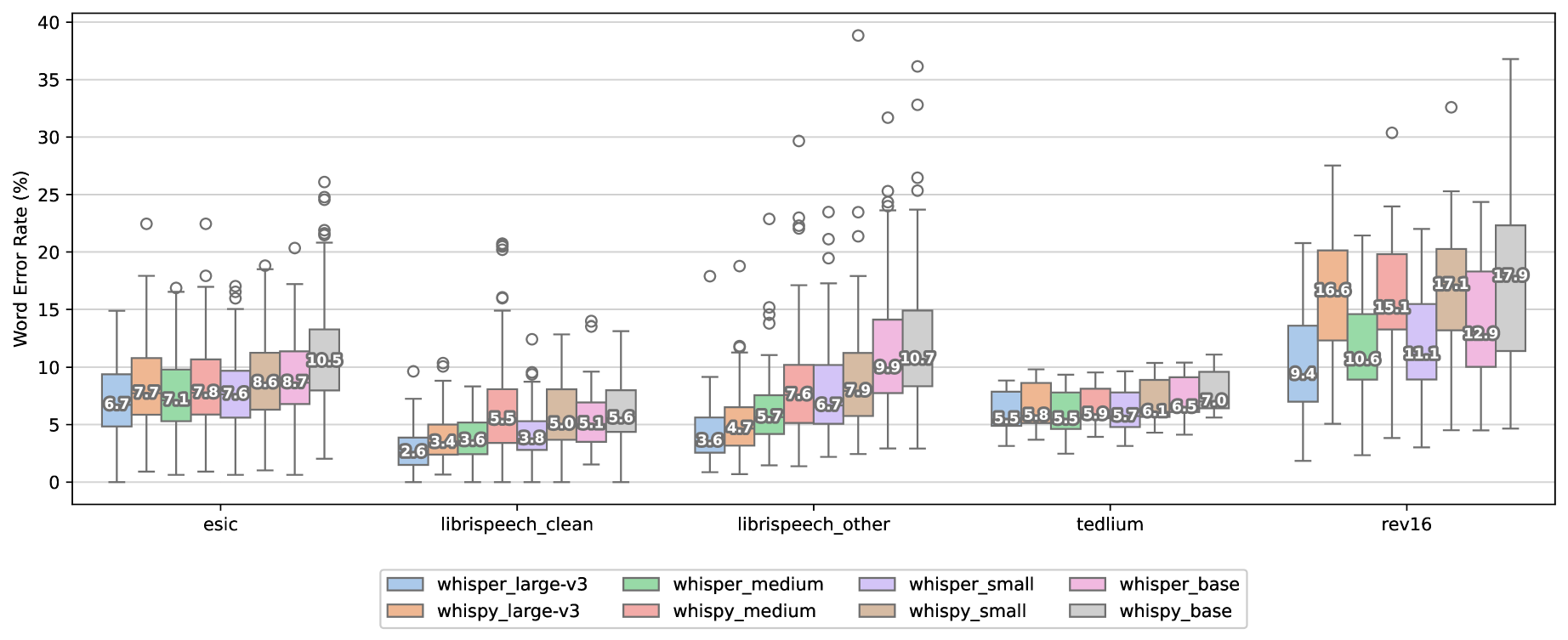}
    \caption{Across all the tested datasets, excluding rev16, each instance of our Whispy implementation performs within a 1-2\% negative difference from its corresponding offline Whisper version. Each box in the graph represents the distribution of the Word Error Rate scored by the labeled model.}
    \label{fig:allResults}
\end{figure}

Figure~\ref{fig:allResults} presents the overall WER distributions for all the tested systems across all the tested datasets. The results show how Whispy falls short of Whisper by only 1\% to 2\% in terms of error rate, with the exception of the rev16 dataset, for which the difference spans from 5\% to 7\%.

\begin{figure}
    \centering
    \includegraphics[width=\textwidth]{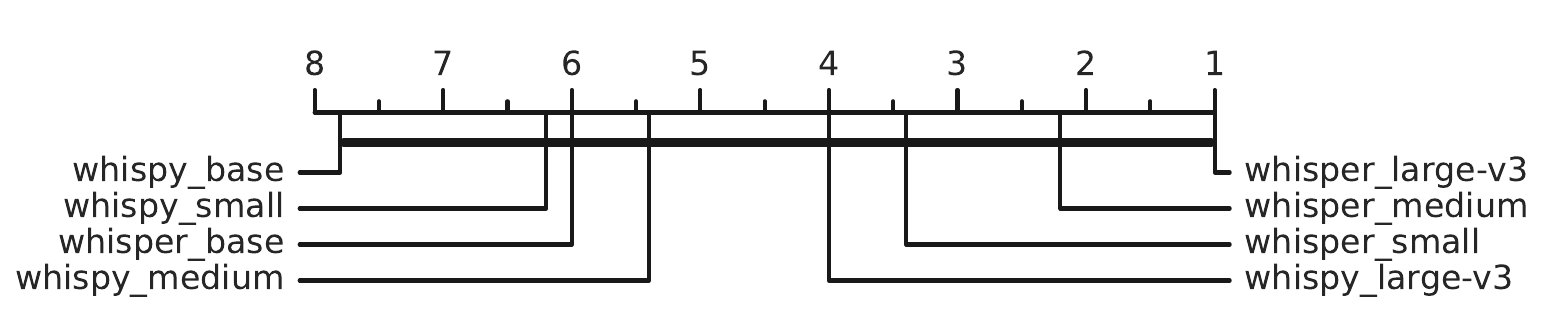}
    \caption{Diagram of the critical differences among the tested models. The continuous bold line suggests there are no statistically significant differences in the obtained results, despite offline Whisper models performing, on average, better than real-time Whispy models.}
    \label{fig:cddiagram}
\end{figure}

To get a better understanding of the performance difference between the tested offline and real-time methods, we additionally present the Critical Difference (CD) diagram of our tests in Figure \ref{fig:cddiagram}. CD diagrams are compact tools used to aggregate the results of multiple statistical significance tests. Each model is ranked based on its score across the datasets, and models for which the pairwise score difference is not statistically significant, are merged through a bold horizontal line. CD diagrams rely on Wilcoxon signed-ranks test \cite{JMLR:v7:demsar06a}. According to our results, Whispy is not statistically different from Whisper, regardless of the rankings of the different models under test. It is worth mentioning that we used Word Accuracy instead of WER when inspecting critical differences, as CD diagrams require a rankable metric where higher is better.

\begin{figure}[h!]
    \centering
    \includegraphics[width=\textwidth]{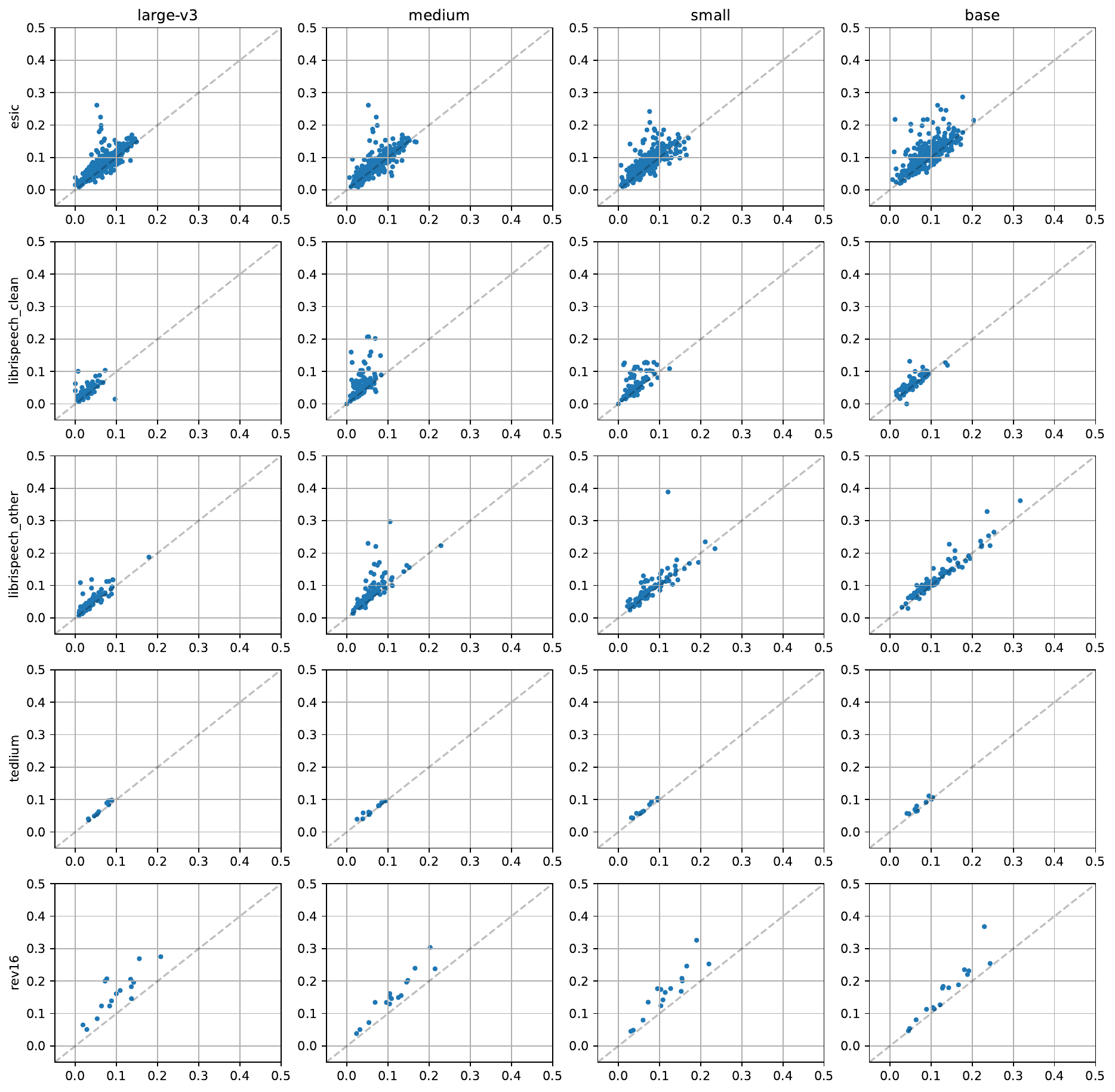}
    \caption{Pairwisewise comparison of offline Whisper transcription WER (x axis) against real-time Whispy transcription WER (y axis). Data points above the quadrant bisector represent audio clips for which Whispy scored lower than Whisper (higher WER), while the region below the quadrant bisector contains all data points for which Whispy scored higher than Whisper (lower WER).}
    \label{fig:metrics}
\end{figure}

Lastly, we inspect the pairwise WER scores for the datasets under test as reported in Figure \ref{fig:metrics}, where each scatter plot shows how Whispy performs against Whisper on a given dataset. In most cases, audio clips cluster around the quadrant bisectors, on which WER scores are identical for both Whisper and Whispy. Some datasets still present a slight upward spray, but the overall tendency is that Whispy is generally aligned with Whisper in terms of WER-based transcription quality.

\subsection{Timing performance}
When evaluating the latency introduced by Whispy, we take into account the temporal delays associated with each one of the steps illustrated in Algorithm \ref{alg:transcription}. These include: VAD processing on the newest chunks (pre-VAD), VAD processing on the entire buffer, actual transcription, and suggestion generation. Hence, we did not focus on the analysis of the overall traversing time for audio chunks received by Whispy, as the decoding delay of the incoming \texttt{ffmpeg} stream is negligible and the study of reception and transmission delays over RTP is out of the scope of this paper. Instead, we collected a set of timers for every transcription chunk produced by Whispy, and then aggregated them into averages for a single audio clip and weighted averages for a full dataset, so that the length of individual clips are considered. Table \ref{table:times} presents a full list of the recorded delays introduced by Whispy, split by dataset and model size. We did not include pre-VAD and VAD times in this list because, due to the adoption of the same Silero pre-trained VAD model regardless of target dataset and Whisper size, they were consistent at 0.06 $\pm$ 0.2 seconds and 0.21 $\pm$ 0.02 seconds respectively across all datasets and model sizes. Similarly, the suggestion generation step required a time always recorded to be around 1 millisecond, so it was excluded from the visualisation. The \textit{total} column, however, accounts for all the available times registered.

\begin{table}[h!]
\caption{Whispy processing times}
\label{table:times}
\centering
\begin{tabular}{ll|ll}
\hline
\textbf{dataset}\hspace{10mm} & \textbf{model}\hspace{10mm}      & \textbf{trx}\hspace{20mm} & \textbf{total}  \\ \hline
                              & large-v3                         & 0.63 $\pm$ 0.07           & 0.88 $\pm$ 0.08 \\
\multirow{2}{*}{ESIC}         & medium                           & 0.63 $\pm$ 0.07           & 0.90 $\pm$ 0.09 \\
                              & small                            & 0.28 $\pm$ 0.03           & 0.55 $\pm$ 0.06 \\
                              & base                             & 0.16 $\pm$ 0.02           & 0.44 $\pm$ 0.05 \\ \hline
                              & large-v3                         & 1.28 $\pm$ 0.14           & 1.56 $\pm$ 0.16 \\
\multirow{2}{*}{libri clean}  & medium                           & 0.88 $\pm$ 0.08           & 1.03 $\pm$ 0.09 \\
                              & small                            & 0.33 $\pm$ 0.03           & 0.61 $\pm$ 0.07 \\
                              & base                             & 0.19 $\pm$ 0.02           & 0.49 $\pm$ 0.05 \\ \hline
                              & large-v3                         & 1.16 $\pm$ 0.11           & 1.44 $\pm$ 0.16 \\
\multirow{2}{*}{libri other}  & medium                           & 0.74 $\pm$ 0.08           & 1.02 $\pm$ 0.11 \\
                              & small                            & 0.33 $\pm$ 0.03           & 0.61 $\pm$ 0.07 \\
                              & base                             & 0.18 $\pm$ 0.02           & 0.46 $\pm$ 0.06 \\ \hline
                              & large-v3                         & 1.24 $\pm$ 0.12           & 1.52 $\pm$ 0.14 \\
\multirow{2}{*}{tedlium}      & medium                           & 0.73 $\pm$ 0.07           & 1.01 $\pm$ 0.09 \\
                              & small                            & 0.32 $\pm$ 0.03           & 0.59 $\pm$ 0.05 \\
                              & base                             & 0.18 $\pm$ 0.02           & 0.45 $\pm$ 0.04 \\ \hline
                              & large-v3                         & 1.39 $\pm$ 0.11           & 1.66 $\pm$ 0.14 \\
\multirow{2}{*}{rev16}        & medium                           & 0.80 $\pm$ 0.06           & 1.06 $\pm$ 0.07 \\
                              & small                            & 0.35 $\pm$ 0.03           & 0.63 $\pm$ 0.05 \\
                              & base                             & 0.20 $\pm$ 0.02           & 0.47 $\pm$ 0.04 \\ \hline
\end{tabular}
\end{table}

Whispy carries a transcription delay spanning from 0.88 seconds to 1.66 seconds when the \texttt{large-v3} Whisper model is used. This delay decreases to a minimum of 0.44 seconds for the \texttt{base} model instances. The low transcription times recorded for the ESIC dataset are caused by the short duration of its audio clips. In fact, when a live transcription starts there is a ramp-up phase when the buffer fills with audio chunks, during which Whispy does not process the entire buffer content but rather a portion of it. In our setup of 5 chunks of 4 seconds each, Whispy first transcribes 4 seconds of audio, then 8 seconds, up until the buffer is full and there is a transcription of 20 seconds of audio every 4 seconds. For audio clips of 1 to 2 minutes, the low processing times at the beginning of the stream favourably skew the average transcription time.

Total delay times should be added to the selected chunk length to obtain the operational delay experienced when transcribing a live audio stream with Whispy. Therefore, in our experimental campaigns transcripts started being produced after roughly 5 seconds of audio activity, and were subsequently delivered at a similar cadence. It is worth mentioning that the Whisper processing times are bound to the choice of hardware and the performance of the selected model, while Whispy introduces a physiological delay caused by its chunking mechanism. However, the length of the chunks acts as an hyperparameter, and can therefore be tuned so that overall latency is reduced.

\subsection{Chunk length and buffer size}
We briefly investigated the effects of configuring Whispy with different values of chunk length and buffer size. The evaluation of these 2 hyperparameters was targeted at the TEDlium dataset only, as it shows stable transcription results in the base experiment line, and it also includes long-form audio clips of a mean duration of around 15 minutes. Tested values for the chunk length are 2, 4, and 6 seconds, while tested values for the buffer size are 3, 5, and 15 chunks.

\begin{figure*}
  \centering
  \subfloat[hyperwer][WER]{
    \includegraphics[width=.4\textwidth]{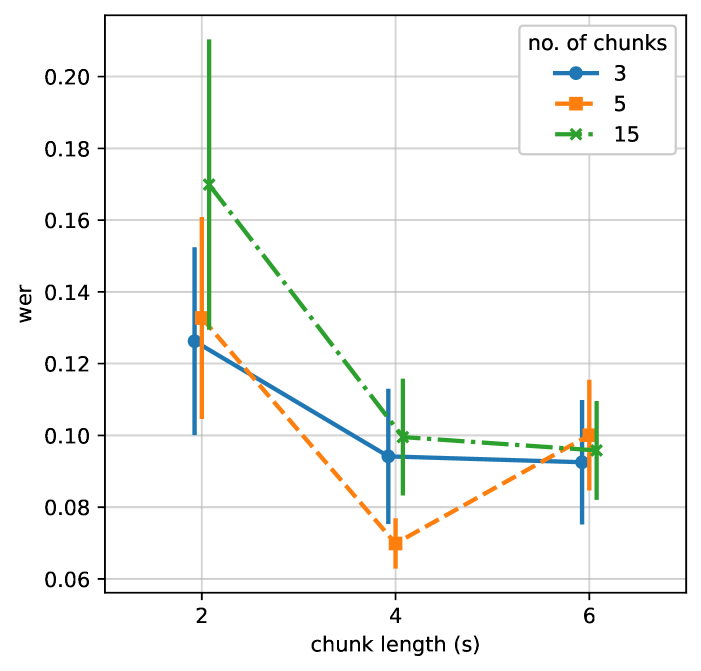}
    \label{hyperwer}
    }
  \qquad
  \subfloat[hyperlatency][Latency]{
    \includegraphics[width=.38\textwidth]{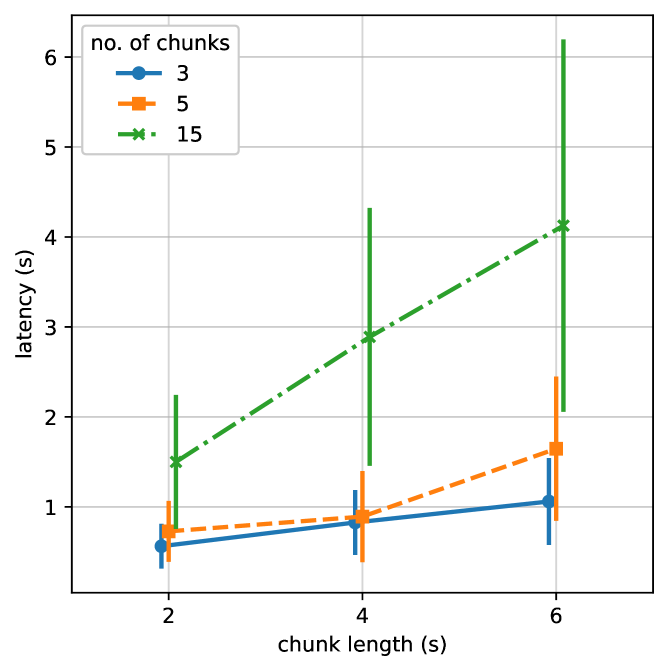}
    \label{hyperlatency}
  }
  \caption{Longer chunks lead to a lower WER, on average, for all the tested models. However, increasing the number of chunks in the buffer worsens the transcription performance (\ref{hyperwer}). The reverse trend appears when measuring the transcription latency, for which combinations of longer chunks and buffer sizes result in longer processing times (\ref{hyperlatency}).}
  \label{pic:hyper}
\end{figure*}

Figure \ref{pic:hyper} shows the breakdown of WER (subfigure \ref{hyperwer}) and transcription latency (subfigure \ref{hyperlatency}) for all the hyperparameter configurations. Error rate in transcriptions generally decreases when longer audio chunks are used, and the opposite behaviour can be seen with respect to the buffer size. While it is certainly immediate to highlight the relationship between the total size of the audio submitted to the transcriber in each iteration and the processing traversal time, as similar delays occur for similar combinations of chunk lengths and buffer sizes, it is less intuitive to pinpoint the reason why different hyperparameters impact the outcome of the suggestions and, therefore, of the transcription quality.

Both chunk size and buffer length impact the Levenshtein distance calculation when a suggestion is generated. Few tokens transcribed differently between two subsequent transcriptions in an overlapping window can vary the distance the same way a small set of new words does. On the other hand, longer chunks are more stable but can introduce larger errors. Small deviations among multiple re-transcriptions will affect the overall transcribed text less than the most recently introduced audio. However, when an error occurs, it can affect a higher number of tokens, thus increasing the WER.

\section{Conclusions and future works}
\label{section:conclusions}

In this paper, we introduced Whispy, an adaptable ASR system built on pre-trained Whisper models, designed for real-time transcription of audio data. Whispy offers remarkable flexibility in deployment and integration with diverse real-world architectures. At its core, Whispy employs a versatile pipeline mechanism, potentially capable of handling multimodal data sources, including video and tabular data. Through rigorous testing on commonly used ASR benchmark datasets, we observed that Whispy maintains transcription performance comparable to its offline counterparts, while exhibiting minimal latency and configurable system settings. Additionally, Whispy features resistance to unexpected source or processing delays.

Moreover, Whispy is slated for production use at the upcoming Internet Engineering Task Force (IETF) meetings. The IETF recordings, along with their corresponding transcriptions generated by Whispy, will provide an invaluable, rich dataset that can be made publicly available to the research community. This dataset holds immense potential for re-training the underlying models and further advancing the state-of-the-art in ASR technology.

Several aspects of Whispy still warrant further exploration and refinement, as ongoing tests may uncover limitations or areas for improvement which will inform future iterations of the system. Firstly, the required steps of VAD can be leveraged to exclude silent regions within the audio buffer, thus reducing the required amount of data processing and lowering the overall system latency. Furthermore, later developments for Whispy may include expanding testing to additional publicly available datasets, increasing the hyperparameter tuning grid for optimization, and enhancing the capabilities of the hallucination filter. Additionally, there is potential to extend Whispy's functionality to encompass tasks such as diarization and summarization, as well as to evolve into a multimodal system capable of processing audio, video, and tabular inputs.

\begin{credits}

\subsection{\discintname}
\subsubsection{\discintname}
The authors are all employed with Meetecho s.r.l. which funded and supported this paper. Meetecho is also the main entity behind the Janus WebRTC server. The authors have otherwise no competing interests to declare that are relevant to the content of this article.
\end{credits}
\bibliographystyle{splncs04}
\bibliography{main}

\end{document}